\documentclass[12pt]{article}
\usepackage{graphicx}
\usepackage{amssymb}
\usepackage{amsmath}
\numberwithin{equation}{section}

\newtheorem{Pa}{Paper}[section]

\newtheorem{Rk}[Pa]{{\bf Remark}}

\newtheorem{Ee}[Pa]{{\bf Example}}
\newtheorem{Dn}[Pa]{{\bf Definition}}

\title{A special structure of the scattering operator \\
and infrared divergences  \\ in quantum electrodynamics}

\author{Lev Sakhnovich}

\date{}

\begin{document}
\maketitle

 E-mail: lsakhnovich@gmail.com

\begin{abstract} We assume that the unperturbed operators $A_0$  are known. Then, the fact that the
scattering operators $S$  and the unperturbed operators $A_0$  are pairwise permutable provides some important information about the structure of the scattering operators. 
Using this information and
the ideas  from  the theory of generalized wave operators,
we present  a new  approach to the divergence
problems in quantum electrodynamics.
We show that the  so called infrared divergences appeared  because
the deviations of the initial and  final waves from the free waves were
not taken into account.\end{abstract}


\vspace{1em}


 \section{Introduction} \label{intro}
 1. In the momentum space the unperturbed  Dirac equation takes the form (see \cite{AB}, Ch.IV):
 \begin{equation}i\frac{\partial}{\partial{t}}\Phi(q,t)=H(q)\Phi(q,t),\quad q=(q_1,q_2,q_3),\label{1.1}\end{equation}
 where $H(q)$ and $\Phi(q,t)$ are matrix functions of order $4{\times}4$ and  $4{\times}1$
respectively. Here the matrix $H(q)$ is defined by the relation \eqref{2.1}.
The unperturbed  Dirac equation in the presence of electro dynamics field
takes the form (see \cite{AB}, Ch.IV):
 \begin{equation}i\frac{\partial}{\partial{t}}\widetilde{\Phi}(q,t)=
 \widetilde{H}(q)\widetilde{\Phi}(q,t),\quad , \label{1.2}\end{equation}
where  $\widetilde{\Phi}(q,t)$ is the  matrix functions of order $8{\times}1$ and
\begin{equation}\widetilde{H}(q)= \left(
                   \begin{array}{cc}
                     H(q) & 0 \\
                     0 & H(q) \\
                   \end{array}
                 \right).
 \label{1.3}\end{equation}
 The equations \eqref{1.1} and \eqref{1.2} can be rewritten in the forms
\begin{equation}i\frac{\partial}{\partial{t}}\Phi(q,t)=A_{0}\Phi(q,t), \label{1.4}\end{equation}
and
\begin{equation}i\frac{\partial}{\partial{t}}\widetilde{\Phi}(q,t)=
 \widetilde{A}_{0}\widetilde{\Phi}(q,t), \label{1.5}\end{equation}
 where the operators ${A}_{0}$ and $\widetilde{A}_{0}$ are defined by
 the relations
 \begin{equation}A_{0}f(q)=H(q)f(q),\quad \widetilde{A}_{0}\widetilde{f}(q)=\widetilde{H}(q)\widetilde{f}(q).\label{1.6}
 \end{equation}
 Here $f(q){\in}L^{2}_{4}(R^3),\quad \widetilde{f}(q){\in}L^{2}_{8}(R^3)$.
 \begin{Rk}\label{Remark 1.1}We do not take into account the perturbed operators $A$ and  $\widetilde{A}$.
 We investigate only the corresponding scattering operators $S(A,{A} _0)$ and
$S(\widetilde{A},\widetilde{A}_0)$.
So,we partially follow the Heisenberg's S-matrix program.In a addition to this S-program
we assume that the unperturbed operators $A_0$ and $\widetilde{A}_0$ are known. This fact gives us an important information about the structure of the scattering operators.\end{Rk}
It is well known that the scattering operators satisfy the following conditions:\\
1) The operators $S(A,{A} _0)$ and
$S(\widetilde{A},\widetilde{A}_0)$ are unitary operators in the spaces
 $L^{2}_{4}(R^3)$ and $L^{2}_{8}(R^3)$ respectively.\\
2) The commutative relations
\begin{equation}A_0S(A,{A} _0)=S(A,{A} _0)A_0,\quad
 \widetilde{A}_0S(\widetilde{A},\widetilde{A}_0)=S(\widetilde{A},\widetilde{A}_0)\widetilde{A}_0
 \label{1.7}\end{equation}
are  valid.\\
\begin{Rk}\label{Remark 1.2}The conditions 1) and 2) are fulfilled for classical and
generalized scattering operators (see \cite{Kat},\cite{Ros}, \cite{Sakh2}).\end{Rk}
In the section 2 we find the eigenvectors and the eigenvalues  of the matrices
$H(q)$ and $\widetilde{H}(q)$.Hence we reduce  the matrices $H(q)$ and $\widetilde{H}(q)$
to the diagonal forms. It follows from conditions 1) and 2) matrices $S(A,{A} _0)$  and $S(\widetilde{A},\widetilde{A}_0)$ can be reduced to the diagonal forms simultaneously with  $H(q)$ and $\widetilde{H}(q)$ respectively. The diagonal elements $d_{k}(A,A_0),\,(1{\leq}k{\leq}4)$ of $S(A,{A} _0)$ and the
diagonal elements $\widetilde{d}_{k}(\widetilde{A},\widetilde{A}_0),\,(1{\leq}k{\leq}8)$ of $S(\widetilde{A},\widetilde{A} _0)$ are such that
\begin{equation}|d_{k}(A,A_0)|=1,\quad (1{\leq}k{\leq}4)\label{1.8}\end{equation}
\begin{equation}|\widetilde{d}_{k}(\widetilde{A},\widetilde{A}_0)|=1,\quad (1{\leq}k{\leq}8)
\label{1.9}\end{equation}
\emph{Thus, the investigation of the scattering operators is reduced to the scalar case,i.e. to the diagonal elements $d_k$.}\\
Using this fact we present a new approach to the divergence problems in quantum electrodynamics (QED). Let us explain the situation.
 In  QED  the higher order approximations of
matrix elements of the scattering matrix contain integrals which diverge.
We think that these divergences are result of the used scattering matrix representation:
the series by a small parameter e. In the section 3 we try to answer J.R.Oppenheimer question \cite{Opp}:\\
 "Can be procedure be freed of the expansion in e and carried out rigorously?"\\
  We introduce a new representation of the scattering
matrix. We do not remove the divergences. Our aim is to prove
that they are absent. In our approach we essentially use the ideas of the generalized wave
operators theory \cite{Sakh2}, \cite{Sakh8},\cite{BuM}.
The classical theory of infrared divergences just rejects the divergent integrals.In our approach this divergent integrals   get the
physical  sense. We construct the deviation factors $U_{0}(L)$ with the help of these integrals.
The invariant region of integration
$\Omega$ is four dimensional sphere with radius $L$.
 The deviation factors $U_{0}(L)$   characterize a deviation of initial and final  waves
from free waves.\\
 \emph{So, we not only have received exact results in the theory of the infraded divergences, but also have received the new facts about behavior of system when the parameter $L$ is great. These facts, as it seems to us, can be  checked by experiment.}
\section{Spectral properties of the matrices $H(q)$, $S(A,A_0)$
 and  $S(\widetilde{A},\widetilde{A}_0)$} \label{spectral}
Let us consider equation \eqref{1.1}.The corresponding matrix $H(q)$ has the form.
\begin{equation} H(q)=\left[ \begin{array}{cccc}
                          m & 0 & q_3 & q_1-iq_2 \\
                          0 & m & q_1+iq_2  & -q_3 \\
                          q_3 & q_1-iq_2  & -m & 0 \\
                          q_1+iq_2  & -q_3 & 0 & -m
                        \end{array}\right] .
\label{2.1}\end{equation}
The eigenvalues $\lambda_{k}$ and the corresponding eigenvectors
$g_k$ of  $H(q)$ are important, and we find them below:
\begin{equation}
\lambda_{1,2}=-\sqrt{m^2+|q|^{2}},\quad \lambda_{3,4}=\sqrt{m^2+|q|^{2}}
\quad ( |q|^2:=q_{1}^2+q_{2}^2+q_{3}^2);
\label{2.2}\end{equation}
\begin{equation}
g_1=\begin{bmatrix}(-q_1+iq_2)/(m+\lambda_3) \\ q_3/(m+\lambda_3) \\ 0 \\ 1\end{bmatrix}, \quad
g_2=\begin{bmatrix}-q_3/(m+\lambda_3) \\ (-q_1- iq_2)/(m+\lambda_3) \\1 \\ 0 \end{bmatrix},
\label{2.3}
\end{equation}
\begin{equation}g_3=\begin{bmatrix} (-q_1+iq_2)/(m-\lambda_3)\\ q_3/(m-\lambda_3)\\ 0 \\ 1\end{bmatrix},
\quad
g_4= \begin{bmatrix} -q_3/(m-\lambda_3) \\ (-q_1-iq_2)/(m-\lambda_3) \\ 1 \\ 0\end{bmatrix}.
\label{2.4}\end{equation}
We introduce the following linear spans:
\begin{equation}M_1(q)=Span\{g_k(q),\,k=1,2\},\,M_2(q)=Span\{g_k(q),\,k=3,4\}.\label{2.5}
\end{equation}
According to condition 2) the subspaces $M_1(q)$ and $M_2(q)$ are invariant subspaces of
$H(q)$ and $S(A,A_0)$. Then there exist common eigenvectors $h_k(q)$ of $H(q)$ and $S(A,A_0)$ such that $h_k(q){\in}M_1(q),\, (k=1,2)$ and $h_k(q){\in}M_2(q),\, (k=3,4).$
Hence $H(q)$ and $S(A,A_0)$ can be reduced to the diagonal forms simultaneously.
In the same way we prove that $\widetilde{H}(q)$ and $S(\widetilde{A},\widetilde{A}_0)$ can be reduced to the diagonal forms simultaneously.
\begin{Rk}\label{Remark 2.1}Article \cite{BMS} contains  formulas, which are similar to 
\eqref{2.2}-\eqref{2.5}  \end{Rk}
\section{New approach to the divergence problems: power series}\label{seqnew}
1. Let the diagonal  element $d(q)$ of the  scattering matrix either $S(A,A_0)$ or $S(\widetilde{A},\widetilde{A}_0)$ be represented in the form of the  power series
\begin{equation}d(q)=1+{\epsilon}a_{1}(q)+{\epsilon}^{2}a_{2}(q)+...
\label{3.1}\end{equation}
We assume that
\begin{equation}
a_{2}=\lim_{L{\to}\infty}\int_{\Omega}F(P,Q)d^{4}P.\label{3.2}\end{equation}
Here $P=[-ip_0,p_1,p_2,p_3],$, $Q=[-iq_0,q_1,q_2,q_3].$ In a number of concrete examples
the functions  $F(P,Q)$ are rational  \cite{AB} . The invariant region of integration
$\Omega$ is four dimensional sphere with radius $L$.\\
We shall investigate the cases when the limit in the right hand side of \eqref{3.2}
does not exist.
\begin{Ee}\label{Example 3.1} Let the relation \begin{equation}a_{2}(q,L)=\int_{\Omega}F(p,q)d^{4}p=i[\phi(q)ln{L}+\psi(q)+O(1/L)],
\,L{\to}+\infty.\label{3.3}\end{equation}
is valid.
Here $\phi(q)=\overline{\phi(q)},\,\psi(q)=\overline{\psi(q)}.$ \end{Ee}
Thus, the corresponding integral (see \eqref{3.3}) diverges logarithmic. Hence the second term of power series \eqref{3.1} is equal to infinity.\\
\emph{Let us
 use a new representation of $d(q)$.}\\
To do it we introduce $d(q,L)$:
\begin{equation}d(q,L)=1+{\epsilon}a_{1}(q)+{\epsilon}^{2}a_{2}(q,L)+...
\label{3.4}\end{equation}We write
\begin{equation}d(q,L)=
L^{i{\epsilon}^{2}\phi(q)}\tilde{d}(q,L)],
\label{3.5}\end{equation}where
\begin{equation}\tilde{d}(q,L)=
[L^{-i{\epsilon}^{2}\phi(q)}{d}(q,L)],
\label{3.6}\end{equation}
Using \eqref{3.4} and \eqref{3.6} we have
\begin{equation}\tilde{d}(q,L)=1+{\epsilon}a_{1}(q)+{\epsilon}^{2}
[a_{2}(q,L)-i\phi(q)ln{L}]+...
\label{3.7}\end{equation}
It follows from \eqref{3.3}  that the second term
\begin{equation}\tilde{a}_{2}(q.L)=a_{2}(q,L)-i\phi(q)ln{L} \label{3.8}\end{equation}
of power series \eqref{3.7} converges when $L{\to}\infty$.
\begin{Rk}\label{Remark 3.2}
The factor $U_{0}(L,q)=L^{i{\epsilon}^{2}\phi(q)}$ is an analogue of deviation factor $W_{0}(t)$ in the theory
of generalized wave and scattering operators \cite{Sakh10}.\end{Rk}
We stress that
\begin{equation}|U_{0}(L,q)|=1.\label{3.9}\end{equation}
\begin{Rk}\label{Remark 3.3}It is  well known (see \cite{AB}, sections 46 and 47 ) that many concrete problems
of collision of particles satisfy the condition \eqref{3.3}.\end{Rk}
\begin{Ee}\label{Example 3.4} Let the relation \begin{equation}a_{2}(q,L)=i[\phi(q)L^{2}+\psi(q)L+{\nu}(q)ln{L}
+\mu(q)+O(1/L)],\label{3.10}\end{equation}
is valid.
Here $\phi(q)=\overline{\phi(q)},\,\psi(q)=\overline{\psi(q)},\,\nu(q)=\overline{\nu(q)},\, \mu(q)=\overline{\mu(q)}$ and $L{\to}+\infty.$\end{Ee}
In this case the factor  $U_{0}(L,q)$ has the form
\begin{equation} U_{0}(L,q)=e^{i{\epsilon}^{2}[\phi(q)L^{2}+\psi(q)L]}L^{i{\epsilon}^{2}\nu(q)}.
\label{3.11}\end{equation} We use \eqref{3.4} and write the formulas
\begin{equation}d(q,L)=U_{0}(L,q)
[\tilde{d}(q,L)],
\label{3.12}\end{equation}where
\begin{equation}\tilde{d}(q,L)=
[U_{0}^{-1}(L,q){d}(q,L)].
\label{3.13}\end{equation}
Relations \eqref{3.7}  in case \eqref{3.10} takes the forms
\begin{equation}\tilde{d}(q,L)=1+{\epsilon}a_{1}(q)+{\epsilon}^{2}
\tilde{a}_{2}(q,L)+...,
\label{3.14}\end{equation}
where term
\begin{equation}\tilde{a}_{2}(q,L)=a_{2}(q,L)-i[\phi(q)L^{2}+\psi(q)L+{\nu}(q)ln{L}] \label{3.15}\end{equation}
of power series \eqref{3.14} converges when $L{\to}\infty$.\\
Relation \eqref{3.9} holds for example 3.4 too.
\begin{Rk}\label{Remark 3.5}All divergences in irreducible diagrams belong
to the class \eqref{3.10}
(see \cite{AB},sections 46 and 47).\end{Rk}
The simplest case of Example 3.4 we obtain when
\begin{equation}\phi(q)=0,\,\nu(q)=0,\,\psi(q)=1.\label{3.16}\end{equation}
In this case we have
\begin{equation}U_{0}(L,q)=e^{i\epsilon^{2}L}.\label{3.17}\end{equation}
2. Now  we assume that
the coefficients $a_{m}(q,L)$ has the form
\begin{equation}a_{m}(q,L)=\sum_{p=0}^{m}[\phi_{p,m}(q)ln^{p}{L}+O(1/L)],\quad
L{\to}\infty,\quad (1{\leq}m{\leq}N).\label{3.18}\end{equation}
It is proved (see review \cite{CD}), that in many cases the Feynman amplitudes
have the poly-logarithmic  structure \eqref{3.18}.
The integrals $a_{m}(q,L)$ which corresponds to the terms of series $a_{m}(q),\,
(1{\leq}m{\leq}N)$ diverges. It was proved in the paper \cite{Sakh3}  that the corresponding deviation factor $U_{0}(L,q)$ has the form
\begin{equation}U_{0}(L,q)=exp[i\sum_{p=1}^{p=N}({\ln}^{p}L)\phi(q,p,\epsilon)], \label{3.19}
\end{equation}
where
\begin{equation} \phi(q,p,\epsilon)=\sum_{m=2}^{m=N}{\epsilon}^{m}\psi(q,p,m),\quad
\psi(q,p,m)=\overline{\psi(q,p,m)}.\label{3.20}\end{equation}
Let us consider the interesting model example.
\begin{Ee}\label{Example 3.6} We assume that the terms $a_{m}(q,L)$ of series
\eqref{2.4} are given by formulas
\begin{equation}a_{m}(q,L)=\sum_{k=0}^{m}\psi_{m-k}(q)\frac{(i\phi(q)lnL)^k}{k!}.
\label{3.21}\end{equation} \end{Ee}
It is easy to see that
\begin{equation}\widetilde{d}(q.L)=L^{-i\epsilon\phi(q)}d(q,L)=1+\epsilon\psi_1(q)+
{\epsilon}^2\psi_2(q)+...\label{3.22}\end{equation}
So, in this case we obtain the regular scattering function $\widetilde{d}(q.L)$.
We note that Coulomb potentials (see \cite{Sakh1} and \cite{Sakh2}) have the properties of type \eqref{3.21}.\\
\begin{Rk}\label{Remark 3.7} Deviation factors are not uniquely defined. If $U_{0}(L,q)$ is the deviation factor.
then $C(q)U_{0}(L,q),\, (|C(q)|=1)$ is  the deviation factor too.
The choice of multipliers $C(q)$ depends on the particular physical problem
under consideration.\end{Rk}
3. Now we introduce the following notion.
\begin{Dn}\label{Definition 3.8} We say say that the deviation factor $U_{0}(L,q)$ belongs to the class
$\mathcal{A}$ if
\begin{equation}U_{0}(L+L_0,q)U_{0}^{-1}(L,q){\to}1,\, L{\to}+\infty.\label{3.23}\end{equation}\end{Dn}
Let us compare the introduced deviation factor $U_{0}(L,q)$ with deviation factor
$W_{0}(t)$ of the generalized wave operators theory \cite{Sakh10}.  The relation \eqref{3.23} for $W_{0}(t)$
has the form
\begin{equation}W_{0}(t+\tau)W_{0}^{-1}(t){\to}1,\, t{\to}\pm\infty.\label{3.24}\end{equation}
\begin{Ee}\label{Example 3.9} If the relation  \eqref{3.18} holds, then the coresponding deviation factors $U_{0}(L,q)$ belong to the class $\mathcal{A}$.\end{Ee}
{\bf Acknowledgements.}
The author is grateful to  A. Sakhnovich and I.~Roitberg for fruitful discussions
and help in the preparation of the manuscript, and to I. Tydniouk for his help with
some important calculations in Section~\ref{spectral}.

\emph{99 Cove ave,  Milford, CT, 06461, USA}

\end{document}